%% file: 0.main.tex
\documentclass[lettersize,journal]{IEEEtran}

\usepackage[ruled,vlined]{algorithm2e}
\usepackage{amsmath}
\usepackage[caption=false,font=normalsize,labelfont=sf,textfont=sf]{subfig}
\DeclareSubrefFormat{myparens}{#1(#2)}
\usepackage{tabularx}
\usepackage{booktabs}
\usepackage{multirow}
\usepackage{makecell}
\usepackage[normalem]{ulem}
\newcommand{\doubleunderline}[1]{\underline{\underline{#1}}}

\usepackage{amsmath,amsfonts}
\usepackage{array}
\usepackage[caption=false,font=normalsize,labelfont=sf,textfont=sf]{subfig}
\usepackage{textcomp}
\usepackage{stfloats}
\usepackage{url}
\usepackage{verbatim}
\usepackage{graphicx}
\usepackage{cite}
\begin{document}

\title{Multi-Scale Diffusion Transformer for Jointly Simulating User Mobility and Mobile Traffic Pattern}

\author{Ziyi Liu, Qingyue Long, Zhiwen Xue, Huandong Wang,~\IEEEmembership{Member,~IEEE}, Yong Li,~\IEEEmembership{Senior Member,~IEEE}%

\thanks{Ziyi Liu, Qingyue Long, Huandong Wang, and Yong Li are with the Department of Electronic Engineering, Beijing
National Research Center for Information Science and Technology (BNRist), Tsinghua University, Beijing 100084, China. (e-mail: liuziyi24@mails.tsinghua.edu.cn; longqy21@mails.tsinghua.edu.cn; wanghuandong@tsinghua.edu.cn; liyong07@tsinghua.edu.cn)}%
\thanks{Zhiwen Xue is with the International School, Beijing University of Posts and Telecommunications, Beijing 100876, China. (e-mail:xzwwinner@bupt.edu.cn)}
}

\maketitle

\begin{abstract}
User mobility trajectory and mobile traffic data are essential for a wide spectrum of applications including urban planning, network optimization, and emergency management. However, large-scale and fine-grained mobility data remains difficult to obtain due to privacy concerns and collection costs, making it essential to simulate realistic mobility and traffic patterns. User trajectories and mobile traffic are fundamentally coupled, reflecting both physical mobility and cyber behavior in urban environments. Despite this strong interdependence, existing studies often model them separately, limiting the ability to capture cross-modal dynamics. Therefore, a unified framework is crucial. In this paper, we propose MSTDiff, a Multi-Scale Diffusion Transformer for joint simulation of mobile traffic and user trajectories. First, MSTDiff applies discrete wavelet transforms for multi-resolution traffic decomposition. Second, it uses a hybrid denoising network to process continuous traffic volumes and discrete location sequences. A transition mechanism based on urban knowledge graph embedding similarity is designed to guide semantically informed trajectory generation. Finally, a multi-scale Transformer with cross-attention captures dependencies between trajectories and traffic. Experiments show that MSTDiff surpasses state-of-the-art baselines in traffic and trajectory generation tasks, reducing Jensen-Shannon divergence (JSD) across key statistical metrics by up to 17.38\% for traffic generation, and by an average of 39.53\% for trajectory generation. The source code is available at:\url{https://github.com/tsinghua-fib-lab/MSTDiff}.  
\end{abstract}


\input{main/1.Introduction}
\input{main/2.Preliminaries}

\input{main/3.Methods}

\input{main/4.Experiments}
\input{main/5.RelatedWork}
\input{main/6.Conclusion}

\bibliographystyle{IEEEtran}
\bibliography{main/reference}

\end{document}

%% file: main/1.Introduction.tex
\section{Introduction}
The widespread deployment of cellular networks and the proliferation of smartphones and IoT devices have greatly increased the scale of mobility data. This growth has accelerated mobile data mining, yielding valuable applications in intelligent transportation, urban planning, and network optimization~\cite{ramamohanarao2017traffic,24OpenData}.

However, obtaining large-scale and fine-grained mobility data remains a significant challenge due to privacy concerns, data collection costs, and regulatory constraints. This limitation hinders the progress of many downstream tasks such as trajectory prediction, traffic forecasting, and network resource management. In this context, realistic and controllable data generation methods become increasingly important. Notably, user trajectories and mobile traffic are inherently coupled, capturing both the physical mobility and cyber behavior of individuals. Modeling their joint distribution is critical to support realistic simulation, forecasting, and robust decision-making in mobility-aware systems. However, existing studies often treat trajectory and traffic generation as separate tasks, lacking a unified framework that captures their interdependence.

Recently, diffusion models have achieved state-of-the-art performance across various generative tasks, including image and video generation~\cite{22image1,22image2,22VideoDiff} and time series modeling~\cite{21CSDI}, owing to their stability and sample quality. In particular, integrating diffusion models with Transformer architectures~\cite{23DiT} enables efficient modeling of long-range temporal dependencies, making them well-suited for complex spatiotemporal data generation tasks.

Despite the potential of generative modeling, joint simulation of user trajectories and traffic remains challenging in the following three aspects:
\begin{itemize}
\item \textbf{Mobile user traffic exhibits periodic and aperiodic characteristics.}
Mobile user traffic is highly stochastic, with substantial variation across individuals. While users often follow daily routines that result in periodic data usage patterns, their mobile traffic is also shaped by individual-level randomness and external events. These irregularities differ widely in both their duration and intensity, leading to bursts of traffic usage at varying temporal scales. This makes it particularly challenging to capture the coexisting periodic and aperiodic dynamics.

\item \textbf{Heterogeneous modeling for mobile user traffic and trajectory.}  
Previous work on diffusion-based trajectory generation has primarily focused on continuous GPS coordinates~\cite{23DiffTraj, 24ConTraj}, thereby overlooking the semantic information of locations, which can be categorized into functional types, with transitions between them reflecting meaningful user behavioral patterns. Applying Gaussian noise in continuous space risks distorting these semantic categories. In contrast, traffic data is inherently a continuous temporal sequence. This discrepancy raises challenges in designing a unified generative model to align and jointly sample across discrete and continuous modalities.

\item \textbf{Complex interactions between mobile user traffic and trajectory.}
Mobile user traffic and trajectories are closely related, with both static and dynamic dependencies. On the one hand, the volume of data usage is influenced by geographic context, as the types of apps used often vary depending on the user’s physical location. As users move across different regions, their data consumption changes accordingly. On the other hand, even when users remain at the same location, their traffic patterns can fluctuate significantly due to switching between different apps within short time intervals. This occurs at a finer temporal resolution than user movement, making it challenging to capture dependencies between traffic data and trajectories. 

\end{itemize}

In this work, we propose \textbf{MSTDiff}, a \underline{M}ulti- \underline{S}cale \underline{T}ransformer-based  \underline{Diff}usion model for jointly generating user mobile traffic and user trajectories. To address the first challenge, we apply discrete wavelet transforms to the traffic data to enable multi-resolution modeling and representing, allowing the model to capture bursty patterns at varying temporal scales. Building on this decomposition, we introduce a multi-scale Transformer based on the cross-attention mechanism to capture the temporal correlations between traffic and trajectory across different resolutions. Finally, we develop a hybrid denoising framework to jointly model the denoising processes for both continuous traffic and discrete trajectory data. Specifically, we incorporate urban knowledge graph embeddings to encode the semantic information of locations and design a similarity-based transition matrix that guides the discrete diffusion process.

In conclusion, the main contributions of our work are summarized as follows:
\begin{itemize}
\item We propose a unified diffusion-based framework that jointly generates continuous traffic data and discrete trajectory sequences.
\item We employ wavelet transforms and multi-scale attention mechanisms to model interactions between traffic and trajectories at different resolutions, and design a knowledge-guided discrete diffusion process for trajectory generation.
\item We conduct experiments on a large scale real world mobility dataset and show that our model achieves up to 17.38\% improvement in Jensen–Shannon divergence on traffic data and an average improvement of 39.53\% on the trajectory task.
\end{itemize}

The conference version of this work is accepted as a short paper at SIGSPATIAL 2025~\cite{liu2025traffic}. In this paper we add trajectory generation experiments with results and visualizations of evaluation statistics, update the loss by adding a cross entropy prediction term for the trajectory branch, and introduce the training and sampling algorithms for the co-denoising process.

%% file: main/2.Preliminaries.tex
\section{Preliminaries and Problem Definition}

\subsection{Continuous Diffusion Models}
Continuous diffusion models are particularly well-suited for modeling real-valued data, such as the temporal sequences. Based on the forward and reverse processes, diffusion models~\cite{DDPM} simulate data distribution by first perturbing clean data with noise and then learning to reverse this corruption to recover the original distribution. For an input space \( X \subseteq \mathbb{R}^D \),  we consider a data point \( x_0 \in X \) sampled from a distribution \( q(x_0) \). The forward process produces a sequence of noisy variables \( \{x_1, x_2, \ldots, x_S\} \) by gradually perturbing \( x_0 \) with noise. The model is trained to learn a distribution \( p_\theta(x_0) \) that closely resembles \( q(x_0) \).


\textbf{Forward process}.
In the forward process, noise is gradually added to the data step by step. This process follows a Markov chain and is defined as:
\begin{equation}
q(x_s \mid x_{s-1}) = \mathcal{N}(\sqrt{1 - \beta_s}x_{s-1}, \beta_s \mathbf{I}),
\end{equation}
where \(\beta_s\) is a small number that slowly increases with each step \(s\), meaning the noise becomes stronger over time and less initial information is retained. Define \(\alpha_s = 1 - \beta_s\) and \(\overline{\alpha}_s = \prod_{i=1}^s \alpha_i\). Then, the noisy sample \(x_s\) can be written as: \(x_s = \sqrt{\overline{\alpha}_s} x_0 + \sqrt{1 - \overline{\alpha}_s} \epsilon\), where \(\epsilon \sim \mathcal{N}(0, \mathbf{I})\).

\textbf{Reverse process}.
The reverse process begins from a fully noisy input \(x_S \sim \mathcal{N}(0, \mathbf{I})\) and aims to progressively recover the clean data. The denoising step is modeled as:
\begin{equation}
p_\theta(x_{s-1} \mid x_s) = \mathcal{N}(x_{s-1}; \mu_\theta(x_s, s), \sigma_\theta(x_s, s)\mathbf{I}).
\label{eq:imp}
\end{equation}

The model parameters \(\theta\) are learned by minimizing a variational upper bound on the negative log-likelihood. Ho et al.~\cite{DDPM} reformulated this optimization into a simpler form, where the objective is to minimize the Mean Squared Error (MSE) between the predicted noise \(\epsilon_\theta(x_s, s)\) and the true noise \(\epsilon\):
\begin{equation}
\min_{\theta} \mathcal{L}(\theta) = \min_{\theta} \mathbb{E}_{x_0 \sim q(x_0), \epsilon \sim \mathcal{N}(0, \mathbf{I}), s} \left\| \epsilon - \epsilon_{\theta}(x_s, s) \right\|_2^2.
\label{eq:loss}
\end{equation}

\subsection{Discrete Diffusion Models}
Discrete diffusion models~\cite{22Argmax, D3PM} are designed to handle data with inherently discrete structures, such as categorical variables and text tokens. Both continuous and discrete diffusion models involve a forward and a reverse process, but differ in their implementation details. Unlike continuous diffusion models which add Gaussian noise to real-valued data, discrete diffusion models corrupt the data through stochastic transitions between discrete states, which gradually transforms the data toward a uniform distribution over the state space. As discussed by Austin et al.~\cite{D3PM}, several discrete corruption schemes have been proposed, including discrete Gaussian noise, uniform transition probabilities, and similarity-based transition matrices constructed from embedding spaces. In this work, we focus on transition-matrix-based corruption, motivated by the observation that locations and words share similar characteristics. Both are discrete variables that can be represented in embedding spaces that preserve their semantic meanings.

\textbf{Forward process}.
In a discrete diffusion process with a total of $S$ steps, the one-step noise addition at step $s$ is modeled as a single-step Markov transition governed by a transition matrix $Q_s \in \mathbb{R}^{N \times N}$, where $N$ is the number of discrete categories. Each transition probability $q(x_s \mid x_{s-1})$ is a categorical distribution whose parameters are specified by the corresponding row of $Q_s$. Assuming $x_{s-1}$ is represented as a one-hot vector, the forward process can be defined as:
\begin{equation}
q(x_s \mid x_{s-1}) = \mathrm{Cat}\left(x_s;\, \boldsymbol{p} = x_{s-1} Q_s\right),
\label{eq:forward_Q}
\end{equation}
\begin{equation}
q(x_s \mid x_0) = \text{Cat}(x_s ; \boldsymbol{p}=x_0\bar Q_s),
\end{equation}
where $\bar Q_s = Q_1Q_2\cdots Q_s$.

\textbf{Reverse process.} 
The reverse process aims to recover the original data by gradually denoising the corrupted input through a sequence of learned transitions. Instead of directly predicting the distribution \( p_\theta(x_{s-1} \mid x_s) \), the model is trained to approximate the posterior distribution over the original state \( x_0 \), given a noisy input \( x_s \). The neural network in the denoising network outputs a parameterized distribution \( p_\theta(x_0 \mid x_s) \), typically in the form of logits over all discrete categories. The reverse transition \( p_\theta(x_{s-1} \mid x_s) \) can then be computed as follows:
\begin{equation}
p_\theta(x_{s-1} \mid x_s)  = \sum_{\hat x_0}q(x_{s-1}\mid x_s,\hat x_0) p_\theta(\hat x_0\mid x_s).
\end{equation}

To train the denoising model, discrete diffusion models typically minimize the Kullback-Leibler (KL) divergence between the true reverse posterior \( q(x_{s-1} \mid x_s, x_0) \) and the model's predicted reverse transition \( p_\theta(x_{s-1} \mid x_s) \). The loss function is defined as:
\begin{equation}
\text{D}_{KL}(q(x_{s-1}\mid x_s, x_0)|| p_\theta(x_{s-1}\mid x_s).
\end{equation}

\subsection{Problem Definition}
Given a set of mobile user traffic sequences $M \in \mathbb{R}^{U \times T}$ and corresponding trajectory sequences $L \in \{1, \dots, N\}^{U \times T}$, where $U$ denotes the number of users, $T$ is the total number of time steps, and $N$ is the number of possible spatial locations, our goal is to train a unified model $F_\theta$ that jointly simulates both traffic data $\hat{M}$ and discrete trajectories $\hat{L}$. Therefore, the joint generative process is defined as:
\begin{equation}
\hat{M}, \hat{L} = F_\theta(\epsilon_S, l_S).
\end{equation}

%% file: main/3.Methods.tex
\section{Methods}
\begin{figure*}[t!]
\includegraphics[width=0.96\textwidth]{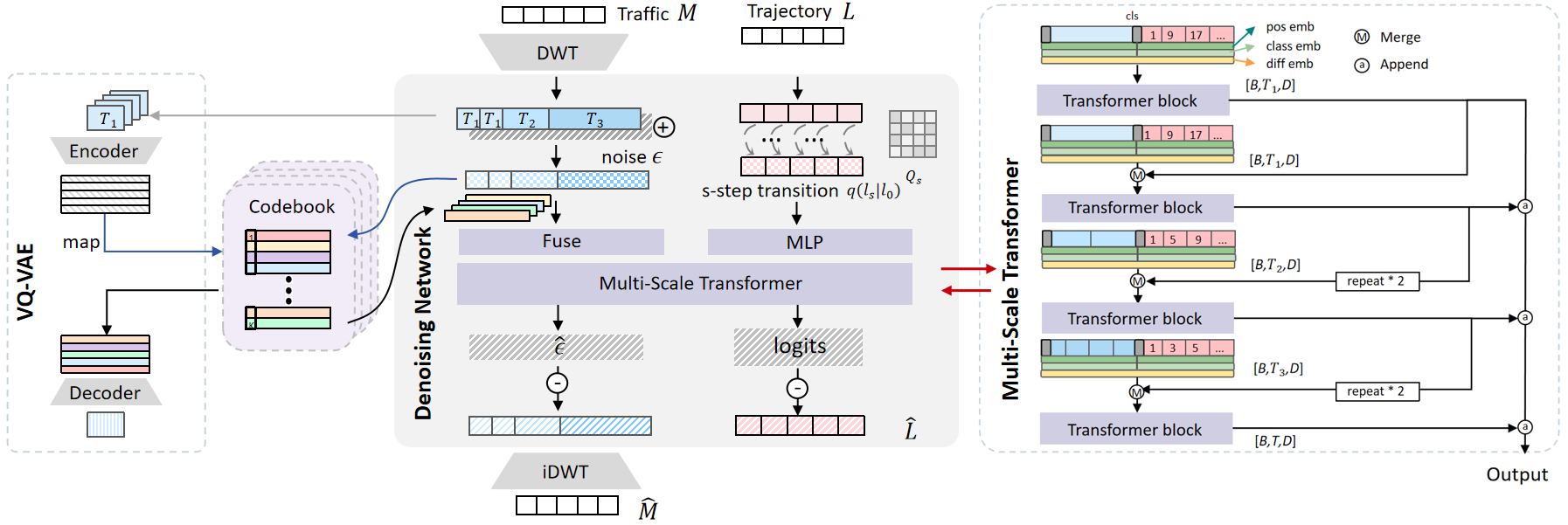}
    \vspace{-0.3cm}
    \caption{The Framework of MSTDiff.}
    \label{fig:framework}
    \vspace{-0.5cm}
\end{figure*}
The framework of MSTDiff is illustrated in Figure~\ref{fig:framework}, which consists of three main modules: a traffic representation module based on discrete wavelet transform and a pre-trained Vector Quantized Variational Autoencoder (VQ-VAE), a trajectory modeling module that constructs a transition mechanism using an urban knowledge graph, and a multi-scale Transformer denoising network for the joint generation of continuous traffic data and discrete trajectories.

\subsection{Wavelet-VQ Module for Traffic Representation}
\subsubsection{Wavelet Transform}
The Discrete Wavelet Transform (DWT) enables multi-resolution analysis through adaptable lengths of basis functions, making it well-suited for traffic sequence that exhibit strong non-stationary and multi-scale temporal dynamics.

Given the original traffic sequence \(m_0\), we apply a three-level DWT to obtain the coefficient sets \(\{CA_3, CD_3, CD_2, CD_1\}\), where \(CA_j\) and \(CD_j\) denote approximation and detail coefficients, respectively. DWT applies low- and high-pass filters at each level with downsampling by a factor of 2. With the Daubechies 1 (db1) wavelet, each level’s coefficients are exactly half the length of the previous, and their total length equals the original sequence length $T$. We then concatenate these coefficients along the last dimension to form a unified multi-scale representation $w_0$.


\subsubsection{Pre-training VQ-VAE}
The VQ-VAE is applied to each coefficient set separately to discretize the wavelet coefficients, enabling the capture of high-level traffic patterns with greater stability than continuous latent representations. The encoder transforms the input traffic sequence $w$ into a continuous latent representation $z_e$. The vector quantization module maps each vector in $z_e$ to its nearest codebook entry from $\mathcal{Z} = \{e_1, e_2, \dots, e_K\}$ to obtain $z_q$. The decoder reconstructs the input sequence. The loss function is defined as:
\begin{equation}
L = \underbrace{\|w - \hat{w}\|_2^2}_{\text{Reconstruction loss}} 
+ \underbrace{\|\text{sg}[z_e] - z_q\|_2^2}_{\text{VQ loss}} 
+ \beta \underbrace{\|z_e - \text{sg}[z_q]\|_2^2}_{\text{Commitment loss}},
\end{equation}
where $\text{sg}[\cdot]$ represents the stop-gradient operator. The first term $\mathcal{L}_{\text{rec}}$ ensures that the reconstructed traffic closely matches the original input. The second term $\mathcal{L}_{\text{VQ}}$ updates the codebook vectors to better align with the encoder outputs by pulling codewords toward the latent representations. The third loss $\mathcal{L}_{\text{com}}$ encourages the encoder outputs to stay close to their assigned codebook entries, preventing the encoder from drifting too far from the discrete latent space.

The pretrained encoder and codebook are then used to extract discrete latent representations of traffic data as inputs to the denoising process.

\subsection{Discrete Diffusion for Trajectory Modeling}
\subsubsection{Urban Knowledge Graph Embedding}
\begin{figure}[t]
\includegraphics[width=0.5\textwidth]{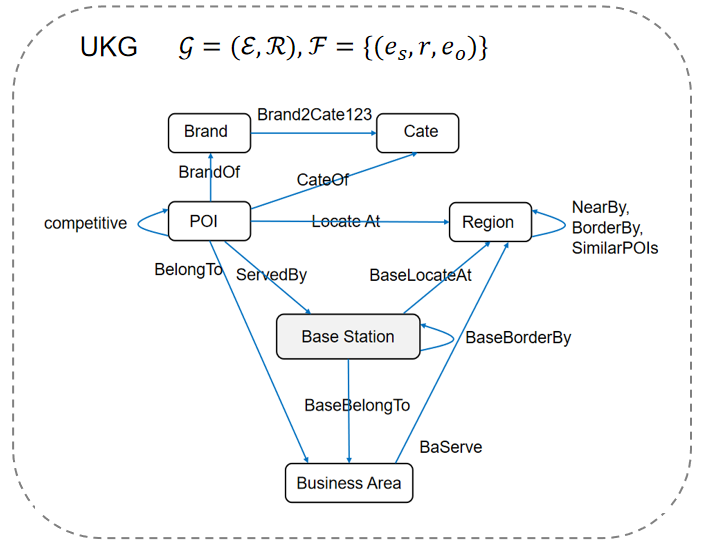}
    \caption{Illustration of urban knowledge graph}
    \label{fig:UKG}
\end{figure}
Prior studies have demonstrated that the semantic characteristics of places can lead to distinct traffic patterns at base stations~\cite{Wang15Traf_pattern, Smartphone, Decom}. These traffic patterns influence the data usage behaviors of mobile users. To effectively capture the semantic context of the urban environment, we leverage an urban knowledge graph, which represents relationships between various urban entities in the form of triplets. The urban knowledge graph $\mathcal{G}$ consists of a set of triplets denoted by $\mathcal{F} = {(e_s, r, e_o)}$, where $e_s$ and $e_o$ are entities in $\mathcal{E}$. Each relation $r \in \mathcal{R}$ encodes a semantic connection between the corresponding entities. The entities and relations are shown in Fig.~\ref{fig:UKG}. Specifically, since base station IDs represent the locations, we focus on relations that are directly associated with base stations:
(1) \textbf{BaseLocateAt}: a base station is located within a region;
(2) \textbf{BaseBelongTo}: a base station is associated with a business area;
(3) \textbf{BaseBorderBy}: a base station shares a border with another base station;
(4) \textbf{ServedBy}: a base station is served by nearby points of interest (POIs).
To learn knowledge graph embeddings for these location entities, we adopt the TuckER model~\cite{TuckER}, which is an effective tensor factorization approach for knowledge graph representation learning.

\subsubsection{Construction of Transition Matrix}
The forward process of a discrete diffusion model, defined in Equation~\ref{eq:forward_Q}, requires the design of a corruption scheme that gradually transforms the data toward a uniform distribution over the state space. Here, we adopt the transition-matrix approach, motivated by the analogy between locations and words, as both are discrete variables whose embeddings preserve semantics. 

After obtaining the output of TuckER, we select the top-$K$ nearest neighbors within a distance threshold for each location embedding to form a binary adjacency matrix $A \in \mathbb{R}^{N \times N}$, where
\begin{equation}
A_{ij} = 
\begin{cases}
1, & \text{if } j \in \text{Nei}(i) \\
0, & \text{otherwise}
\end{cases}.
\end{equation}

Next, we apply normalization to matrix $A$ as $B = \frac{1}{2K}(A + A^\top)$. The transition rate matrix $R$ has rows summing to zero, with $R_{ij}$ denoting the transition rate from location $i$ to $j$:
\vspace{-0.1cm}
\begin{equation}
R_{ij} = 
\begin{cases}
B_{ij}, & \text{if } i\ne j \\
-\sum_{k\ne i}B_{ik}, & \text{if }i=j
\end{cases}.
\end{equation}
\vspace{-0.1cm}

The one-step transition probability matrix $Q_s$ at diffusion step $s$ and the cumulative transition matrix $\bar{Q}_s$ are given by:
\vspace{-0.1cm}
\begin{equation}
Q_s = \exp(\alpha_{tj,s} R), \quad \bar{Q}_s = \prod_{i=1}^{s} Q_i=\exp(\bar{\alpha}_{tj,s} R),
\end{equation}
where $\bar{\alpha}_{tj,s} = \sum_{i \leq s} \alpha_i$. The transition schedule $\alpha_{tj,s}$ over diffusion steps ensures that at the final step the transition probability approximates a uniform distribution, implying that location information becomes fully randomized.

\begin{algorithm}[t]
\caption{Training}
\label{alg:Training}
\For{\( i = 1 \) to \texttt{ITER}}{
$m_0,l_0 \sim q(M_0), q(L_0)$\; 
$w_0=\text{DWT}(m_0)$\;
$h_0=\text{OneHot}(l_0)$\;
$s \sim \text{Uniform}(\{1,...,S\})$\;
$\epsilon \sim \mathcal{N}(0,1)$\;

\textbf{1.Add noise for traffic data:}

$w_s=\sqrt{\bar \alpha_{tr,s}}w_0+\sqrt{1-\bar \alpha_{tr,s}}\epsilon$\;

\textbf{2.Add noise for trajectory:}

$\bar Q_s=\exp\left(\bar \alpha_{tj,s}R\right)$\;
$q(l_s \mid l_0) =  h_0 \cdot \bar Q_s $\;

\textbf{3.Compute loss:}

$\hat{\epsilon}_s = f_{\theta,tr}(w_s, q(l_s\mid l_0), s)$\;
$\mathbf{z}_s = f_{\theta,tj}(w_s, q(l_s\mid l_0), s)$\;
$\mathcal{L} = \text{MSE}(\epsilon, \hat{\epsilon}_s) + D_{\text{KL}}(q(l_{s-1}\mid l_s, l_0) || p_\theta(l_{s-1}\mid l_s)))$+$\text{CE}(l_0, \mathbf{z}_s)$.
}
\end{algorithm}

\begin{algorithm}
\caption{Sampling}
\label{alg:Sampling}
$w_S \sim \mathcal{N}(0,1)$\;
$l_S \sim \text{Uniform}(\{0, 1, ..., N{-}1\})$\;
 
$p_L = \frac{1}{N} \cdot \mathbf{1}_N \in \mathbb{R}^N$\;

\For{\( s = S \) to \texttt{1}}{
\textbf{1.Impute samples for traffic data:}

$z \sim \mathcal{N}(0,1)$\;
$w_{s-1}=\frac{1}{\sqrt{\alpha_{tr,s}}}(w_s-\frac{1-\alpha_{tr,s}}{\sqrt{1-\bar \alpha_{tr,s}}}f_{\theta,tr}(w_s, p_L, s))+\sigma_s z$\;

\textbf{2.Impute samples for trajectory:}

$\mathbf{z}_s=f_{\theta,tj}(w_s, p_L, s)$\;
$p_\theta(\hat l_0\mid l_s)=\text{softmax}(\mathbf{z}_s)$\;
$\hat l_0 \sim p_\theta(\hat l_0 \mid l_s)$\;

\textbf{2.1 Compute posterior:}

$\hat h_0, h_s=\text{OneHot}(\hat l_0), \text{OneHot}(l_s)$\;
$\textbf{Let } \tilde q_s=h_s Q_s^\top,  \quad \tilde q_{s-1}=\hat h_0 \bar Q_{s-1}$\;

$q(l_{s-1}\mid l_s, \hat l_0) = \text{Cat}(l_{s-1};p=\text{Normalize}(\tilde q_s \odot \tilde q_{s-1})$\;

\textbf{2.2 Compute one-step probability:}

$p_\theta(l_{s-1}\mid l_s)=\sum_{\hat l_0}q(l_{s-1}\mid l_s, \hat l_0)p_\theta(\hat l_0\mid l_s)$\;

$p_L \leftarrow p_\theta(l_{s-1}\mid l_s)$\;

$l_{s-1}\sim p_\theta(l_{s-1}\mid l_s)$\;

}
$m_0=\text{IDWT}(w_0)$\;
\textbf{Return $\{m_0, l_0\}$}.
\end{algorithm}

\subsection{Co-Denoising Process}
\subsubsection{Multi-Scale Transformer Denoising Network}
This module captures temporal dependencies between traffic and trajectory data. We obtain noisy DWT coefficients $w_s \in \mathbb{R}^{B\times T \times 1}$ and the transition probability $q(l_s \mid l_0) \in \mathbb{R}^{B\times T \times N}$, where $B$ denotes the batch size. They are concatenated along the last dimension and fed into the multi-scale Transformer $f_{\theta}(w_s, q(l_s \mid l_0), s)$. Starting from the coarsest wavelet scale, traffic is encoded by a pre-trained VQ-VAE and then fused with the original traffic features via a gating mechanism, while the corresponding trajectory representation is processed by an MLP. Positional, type and diffusion step embeddings are added before entering Transformer blocks. To preserve coarse-scale context, each Transformer output is merged with the finer-scale representation as input to the next layer, enabling hierarchical refinement across scales. The final output is formed by concatenating the outputs of all layers from coarse to fine.

\subsubsection{Loss Function}
The model simultaneously takes continuous traffic data and discrete trajectories as input. Although they share the same diffusion step $s$, we adopt separate noise schedules to reflect their distinct data characteristics. The multi-scale Transformer outputs predictions separately: (1) the predicted noise \( \hat \epsilon_s = f_{\theta,tr}(w_s, q(l_s \mid l_0),s) \) for the traffic data, and (2) the predicted logits \( \mathbf{z}_s = f_{\theta,tj}(w_s, q(l_s \mid l_0), s) \) for the trajectory data.

After obtaining $\hat{w_0}$ via denoising, we apply the inverse Discrete Wavelet Transform (iDWT) to reconstruct the original input. The DWT–iDWT process acts as an encoder–decoder structure, mapping the time series to a latent domain for denoising.

For trajectory data, the multi-scale Transformer outputs logits \(\mathbf{z}_s\), and applying softmax yields \(p_{\theta}(\hat{l}_{0}\mid l_{s})=\mathrm{softmax}(\mathbf{z}_s)\). Given this, the reverse transition probability \( p_\theta(l_{s-1} \mid l_s) \) is computed by marginalizing over possible original states. The total training objective consists of an MSE loss for the traffic branch introduced in Equation~\ref{eq:loss}, and a KL divergence loss for the trajectory branch. In addition, we introduce an auxiliary prediction loss for the trajectory branch that directly supervises the conditional distribution \(p_{\theta}(\hat{l}_{0}\mid l_{s})\), leading to more stable training and improved generative quality. Specifically, since \(l_{0}\) is discrete and the Transformer outputs logits, we instantiate this objective as a cross-entropy loss. Given three hyper parameters $\lambda_1$, $\lambda_2$ and $\lambda_3$, the final loss is given by:
\begin{equation}
\begin{aligned}
\mathcal{L} &= \lambda_1 \cdot \mathcal{L}_{tr} + \lambda_2 \cdot \mathcal{L}_{tj}+\lambda_3 \cdot \mathcal{L}_{pred}, \\
\mathcal{L}_{tr} &= \text{MSE}(\epsilon,\hat \epsilon_s), \\
\mathcal{L}_{tj} &= \text{D}_{\text{KL}}(q(l_{s-1} \mid l_s, l_0) \,\|\, p_\theta(l_{s-1} \mid l_s)),\\
\mathcal{L}_{pred} &= \text{CE}(l_0,\mathbf{z}_s).
\end{aligned}
\end{equation}

In conclusion, the co-denoising module allows the simultaneous reconstruction of both modalities while preserving their temporal alignment and cross-modal dependencies. Detailed training and sampling algorithms are shown in Algorithm~\ref{alg:Training} and Algorithm~\ref{alg:Sampling}.





%% file: main/4.Experiments.tex
\section{Experiments}
\subsection{Experiment Settings}
\subsubsection{Dataset and Preprocessing}
\begin{figure}
    \centering
    \subfloat[User normalized traffic]{%
\includegraphics[width=0.25\textwidth]{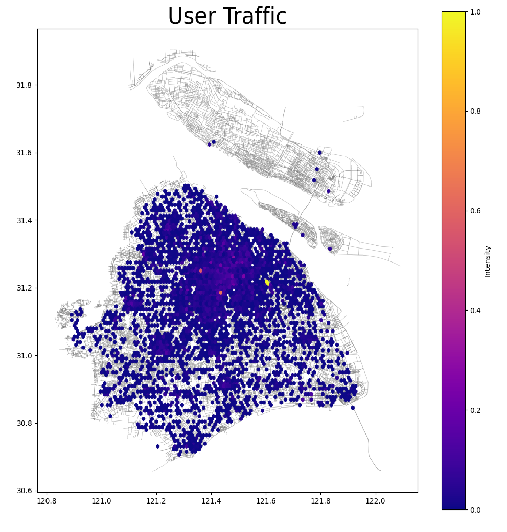}
        \label{fig:data_traf}
    }
    \subfloat[User trajectory]{%
        \includegraphics[width=0.25\textwidth]{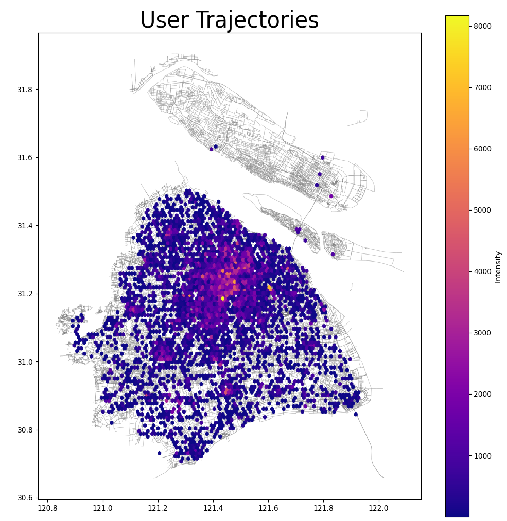}
        \label{fig:data_traj}
    }
    \vspace{-0.2cm}
     \caption{Visualization of dataset.}
    \label{fig:dataset}
\vspace{-0.4cm}
\end{figure}

The large-scale user-level cellular traffic dataset was collected in Shanghai in 2016 and spans one week of data. It captures individual users' mobile network usage over time and space, with each record containing a timestamp, the associated base station ID and its GPS coordinates, and the corresponding traffic volume. Visualizations of the dataset are shown in Figure~\ref{fig:dataset}. After retaining only the city center region, the dataset includes over 2,000 users and approximately 4,000 unique base station locations, with a temporal resolution of 30 minutes, yielding 336 time points per user for the week.

\begin{table*}[t]
\renewcommand{\arraystretch}{1.5}
\centering
\caption{Evaluation results of traffic generation. Bold indicates the best performance, underlined indicates the second-best, and double-underlined indicates the third-best.}
\label{tab:exp-traf}
\begin{tabularx}{0.8\textwidth}{p{2.7cm}|>{\centering\arraybackslash}p{2.7cm}|>{\centering\arraybackslash}p{2.7cm}|>{\centering\arraybackslash}p{2.7cm}}
\toprule
\multirow{2}{*}{\textbf{Methods}} & \multicolumn{1}{c|}{\textbf{Traffic Volume}} & \multicolumn{1}{c|}{\textbf{\makecell{First-order Difference}}} & \multicolumn{1}{c}{\textbf{\makecell{Daily Periodic Component}}} \\ 
\cmidrule(lr){2-4} 
 & \textbf{JSD} & \textbf{JSD} & \textbf{RMSE} \\
\midrule
CSDI & \makecell{\doubleunderline{0.1514}} & \makecell{\underline{0.1669}} & \makecell{\underline{0.0160}} \\

DiT & \makecell{0.2272} & \makecell{0.1889} & \makecell{0.0178} \\

ADAPTIVE & \makecell{\underline{0.1289}} & \makecell{\doubleunderline{0.1888}} & \makecell{\textbf{0.0152}} \\

MSH-GAN & \makecell{0.6701} & \makecell{0.5894} & {\doubleunderline{0.0169}} \\

MSTDiff -w/o Traj & \makecell{0.1682} & \makecell{0.1675} & \makecell{0.0191} \\

MSTDiff & \makecell{\textbf{0.1230}} & \makecell{\textbf{0.1379}} & \makecell{0.0263} \\
\bottomrule
\end{tabularx}
\end{table*}

\subsubsection{Baselines}
MSTDiff is compared with the following four baselines for the traffic generation task:
\begin{itemize}
\item \textbf{CSDI}~\cite{21CSDI}:
CSDI is a diffusion-based model originally developed for probabilistic time series imputation. It utilizes a residual network with a two-dimensional Transformer encoder to capture temporal and feature-wise dependencies separately. This architectural design also enables CSDI to be adapted for time series generation tasks.
\item \textbf{DiT}~\cite{23DiT}: DiT integrates the Transformer architecture into the diffusion process to effectively capture long-range dependencies. The original paper proposes three variants of the Diffusion Transformer, which have shown strong performance in image generation. In our experiment, we adapt DiT to handle time series modeling.
\item \textbf{ADAPTIVE}~\cite{ADAPTIVE}: ADAPTIVE is a city-scale cellular traffic generation framework that leverages transfer learning to synthesize traffic in cities lacking historical data. It aligns base station representations between the target and source cities, and generates target-city traffic using a feature-enhanced GAN.
\item \textbf{MSH-GAN}~\cite{24MSH-GAN}: MSH-GAN is a GAN-based framework for user-level mobile traffic generation. It captures individual level behaviors using BiLSTM and self-attention mechanisms, while using a Switch Mode Generator to capture group-level behaviors. 
\end{itemize}

Moreoever, MSTDiff is compared with the following four baselines for the trajectory generation task:
\begin{itemize}
\item \textbf{TimeGEO}~\cite{jiang2016timegeo}: It generates urban mobility patterns by modeling temporal behavior with the weekly home-based tour count, dwell rate, and burst rate, and by selecting locations via a rank-based exploration and preferential return model that balances visits to new places and returns to familiar ones.
\item \textbf{PateGail}~\cite{wang2023pategail}: It generates mobility trajectories with a privacy-preserving imitation learning framework built on generative adversarial imitation learning to mimic the process of human decision-making.
\item \textbf{DiffTraj}~\cite{23DiffTraj}: It generates GPS trajectory points with a UNet–based diffusion model conditioned on trip attributes. It uses continuous movement features such as distance, duration and average speed, together with discrete factors including departure time slot and start and end region IDs.
\item \textbf{VOLUNTEER}~\cite{long2023practical}: It generates trajectories with a variational framework comprising a user VAE for group-level user distributions and a trajectory VAE for individual mobility patterns. The trajectory VAE decouples travel time and dwell time to generates realistic trajectories.

\end{itemize}

\subsubsection{Metrics}
We evaluate the performance of the traffic generation task using three metrics. The Jensen–Shannon divergence (JSD) of traffic volume distribution, the JSD of first-order differences to assess short-term variation patterns, and the root-mean-square error (RMSE) of daily periodic components extracted via Fourier analysis to measure the alignment of daily cycles. Lower values across these metrics indicate closer alignment between generated and real-world traffic data.

We evaluate the performance of trajectory generation using six metrics. All metrics are computed as the JSD between the generated and real distributions. \textbf{Distance} is the distance between adjacent visited locations for each user. \textbf{Radius} is the radius of gyration of a user's trajectory. \textbf{DistinctLoc} is the fraction of distinct locations per user. \textbf{Duration} is the contiguous stay duration per visit. \textbf{G-rank} is the distribution of visit frequency over the locations with the top 10\% overall visits. \textbf{I-rank} is the distribution of visit frequency over each user's own visited locations.

\subsection{Traffic Generation Task}
The results of traffic generation task are shown in Table~\ref{tab:exp-traf}. MSTDiff achieves superior performance in both the JSD of traffic volume and the JSD of first-order differences. Specifically, it reduces the JSD of traffic volume by 4.58\% compared to ADAPTIVE, and the JSD of first-order differences by 17.38\% compared to the best-performing baseline, CSDI. However, the RMSE is relatively higher, possibly due to the use of wavelet transforms, which are well-suited for capturing sudden bursts but less effective in modeling stable daily periodic patterns. 

We also conduct an ablation study where the model uses only traffic data as input, excluding trajectory information. The performance degrades with an increase in JSD. Compared to the ablated version, our full model incorporating both traffic and trajectory inputs achieves a 26.87\% reduction in the JSD of traffic volume, and a 17.67\% reduction in the JSD of first-order differences. These results highlight the effectiveness of incorporating trajectory information in improving traffic generation performance.

\subsection{Trajectory Generation Task}
\begin{figure*}[t!]
\includegraphics[width=0.96\textwidth]{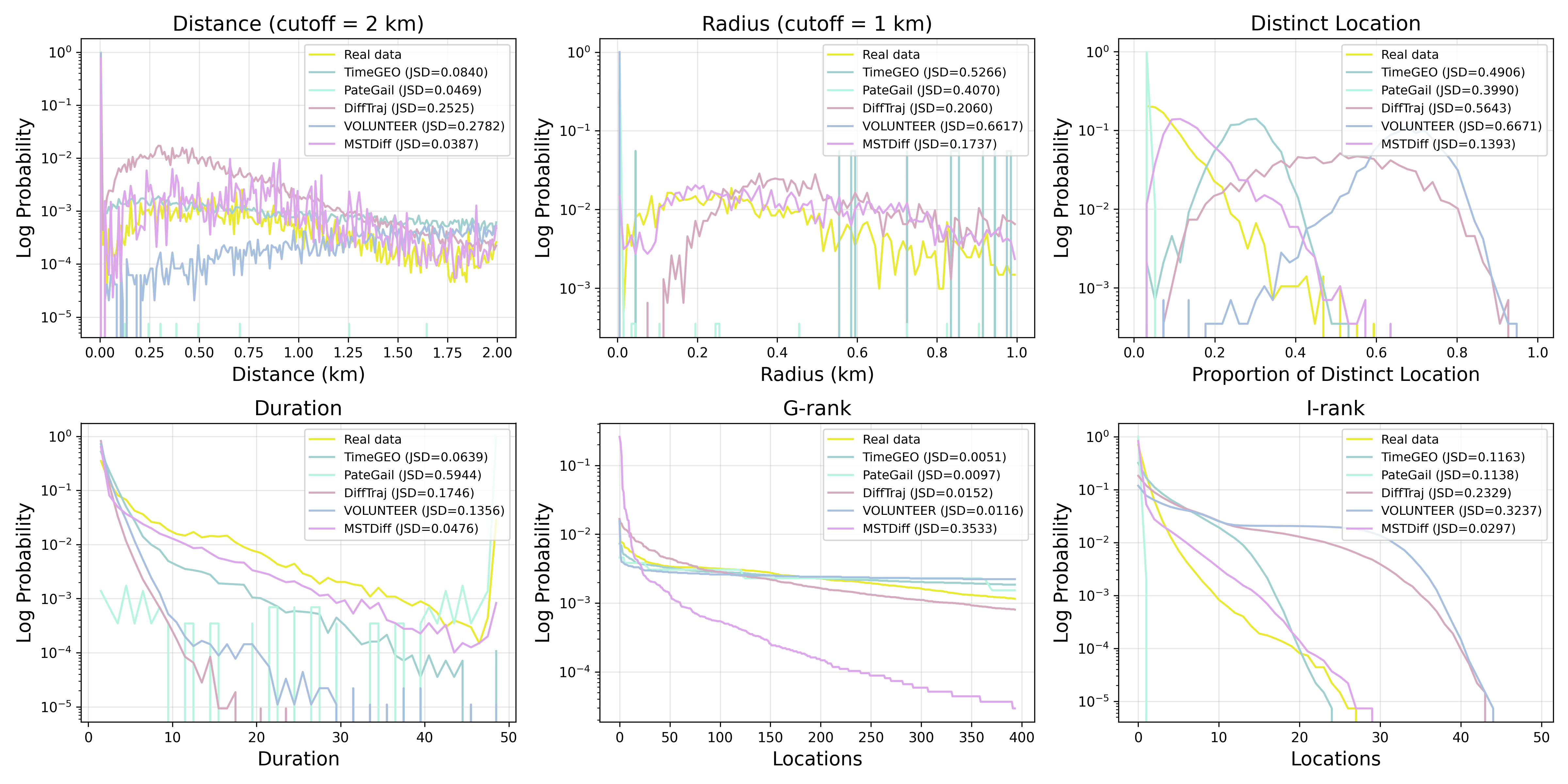}
    \vspace{-0.3cm}
    \caption{Visualization of metrics of trajectory generation.}
    \label{fig:traj-metrics}
    \vspace{-0.5cm}
\end{figure*}

The results of the trajectory generation task are shown in Table~\ref{tab:exp-traj}. MSTDiff achieves the best scores on five of six JSD metrics. Compared with the best performing baselines, it reduces JSD by more than 15\% on Distance and Radius, 65.09\% on DistinctLoc, and 25.51\% on Duration. These gains indicate that the model more faithfully captures users’ location switches, realistic spatial movement, and dwell time distributions, reflecting strong spatiotemporal modeling of trajectories. The distributions of the evaluation statistics are shown in Figure~\ref{fig:traj-metrics}, and in most cases MSTDiff is closest to the real data. In most cases the distribution of MSTDiff is closest to the real data. In contrast, its G-rank performance is weaker than other baseline models, suggesting an overall preference for specific locations, yet it models user-level visiting tendencies effectively. 

\begin{table*}[t]
\renewcommand{\arraystretch}{1.5}
\centering
\caption{Evaluation results of trajectory generation.}
\label{tab:exp-traj}

\begin{tabularx}{0.98\textwidth}{
    p{2.7cm}|
    >{\centering\arraybackslash}p{1.6cm}|
    >{\centering\arraybackslash}p{1.6cm}|
    >{\centering\arraybackslash}p{1.6cm}|
    >{\centering\arraybackslash}p{1.6cm}|
    >{\centering\arraybackslash}p{1.6cm}|
    >{\centering\arraybackslash}p{1.6cm}
}
\toprule
\textbf{Methods} & \textbf{Distance} & \textbf{Radius} & \textbf{DistinctLoc} & \textbf{Duration} & \textbf{G-rank} & \textbf{I-rank} \\
\midrule
TimeGEO              & \doubleunderline{0.0840}
 & 0.5266
 & \doubleunderline{0.4906}
 & \underline{0.0639}
 & \textbf{0.0051}
 & \doubleunderline{0.1163} \\
PateGail          & \underline{0.0469}
 & \doubleunderline{0.4070}
& \underline{0.3990} & 0.5944
 & \underline{0.0097}
 & \underline{0.1138} \\
DiffTraj           & 0.2525
 & \underline{0.2060}
 & 0.5643
 & 0.1746
 & 0.0152
 & 0.2329 \\
VOLUNTEER & 0.2782
 & 0.6617
 & 0.6671
 & \doubleunderline{0.1356}
 & \doubleunderline{0.0116}
 & 0.3237 \\
MSTDiff           & \textbf{0.0387}
 & \textbf{0.1737}
 & \textbf{0.1393}
 & \textbf{0.0476}
 & 0.3533
 & \textbf{0.0297} \\
\bottomrule
\end{tabularx}
\end{table*}

%% file: main/5.RelatedWork.tex
\section{Related work}
\subsection{Trajectory Generation}
Trajectory generation plays a critical role in urban planning, emergency management, and traffic scheduling. Early studies primarily relied on statistical and physical modeling, aiming to uncover universal patterns in human mobility behavior. For example, the Lévy flight model captures the proportion of long- and short-distance movements through a power-law distribution, successfully reproducing the jumpy and irregular characteristics commonly observed in human travel~\cite{brockmann2006scaling}. The Exploration and Preferential Return (EPR) model further introduces location preferences and return mechanisms, offering a more realistic representation of human mobility patterns~\cite{song2010modelling}. The TimeGeo~\cite{jiang2016timegeo} variant of EPR models the relationships between home and work inferred from data and flexible activities' spatial and temporal patterns (e.g., other activities). It sets parameters to capture individuals' circadian rhythms and generates a wide range of empirically observed movement behaviors. 

In recent years, the rapid development of artificial intelligence (AI), especially the rise of deep generative models, has opened up new possibilities for trajectory generation. For example, PateGail leverages powerful generative adversarial imitation learning to simulate human decision-making processes and incorporates privacy-preserving mechanisms to protect user data~\cite{wang2023pategail}. Long et al. proposed the VOLUNTEER framework, which employs a dual VAE architecture to model both user attribute distributions and trajectory behavior distributions, significantly improving the realism and diversity of the generated trajectories~\cite{long2023practical}. Additionally, Zhu et al. introduced a spatiotemporal diffusion probabilistic model for trajectory modeling, effectively combining the diffusion process with the spatiotemporal characteristics of trajectories~\cite{23DiffTraj}. Notably, with the widespread adoption of large language models (LLMs), recent studies have begun to explore their potential in modeling human mobility behavior. Wang et al. proposed an LLM-based agent framework for trajectory generation, enabling the model to understand and generate complex movement trajectories that incorporate individual motivations, preferences, and semantic context~\cite{jiawei2024large}.

However, existing studies focus on the single task of trajectory generation, often overlooking the intrinsic relationship between individual trajectories and user mobile traffic. To address this limitation, we propose a novel framework that jointly generates human mobility trajectories and user mobile traffic.

\subsection{Cellular Traffic Generation}

With the rapid growth of mobile networks, simulating cellular traffic at both the city scale and the individual level has become increasingly important for evaluating network performance and guiding deployment strategies. This section reviews recent studies on cellular traffic generation.

The task of traffic generation has evolved over time. In its early stages, researchers primarily relied on mathematical approaches~\cite{04Harpoon,06Tmix,09Swing} to simulate network traffic. For example, Aceto et al.~\cite{21Markov} used Markov Chains and Hidden Markov Models to model traffic behavior. With the rise of artificial intelligence, more advanced generative models have emerged. Zhang et al.~\cite{18DCNN} introduced a densely connected convolutional network to simulate citywide traffic. LSTM-based models were explored by Dalgkitsis et al.~\cite{18LSTM} and Trinh et al.~\cite{18LSTM2} for time series prediction. More recently, Su et al.~\cite{24Atten} presented a lightweight attention-based deep learning framework to generate weekly cell traffic patterns.

Another line of neural network-based models involves GANs, which have been widely applied to generate both flow-level~\cite{17FlowSyn,19Flow} and packet-level traffic~\cite{19PAC,22Prac,PcapGAN} at individual base stations, and typically rely on detailed network configuration data. As a result, they are not well-suited for large-scale generation tasks. To address this, newer models have adapted GANs for city-level applications~\cite{KE-GAN,ADAPTIVE}. For instance, Hui et al.~\cite{KE-GAN} designed a knowledge-enhanced GAN that incorporates multi-periodic urban features and base station locations. Similarly, Zhang et al.~\cite{ADAPTIVE} proposed ADAPTIVE, a transfer learning framework that aligns traffic patterns from a source city to a target city lacking historical records.

Recently, diffusion models have emerged as strong alternatives in generative tasks, offering greater stability and mitigating issues such as mode collapse that are common in GAN-based approaches. They have been successfully applied in diverse domains, including image generation~\cite{22image1,22image2}, time-series data imputation~\cite{21CSDI}, as well as traffic prediction~\cite{20traffic}. In the field of network traffic generation, Qi et al.~\cite{Qi2024CANDLE} introduced CANDLE, a conditional diffusion model that generates 5G traffic in unobserved regions by learning cross-modal relations between 4G and 5G traffic through attention mechanisms, while using GCNs to model spatial correlations. Another diffusion-based approach, STK-Diff by Chai et al.~\cite{Chai25STK}, integrates modules for urban context, temporal patterns, and spatial relationships to enable realistic and controllable traffic synthesis at city scale. Moreover, a recent model called NetDiffus, proposed by Sivaroopan et al.~\cite{24NetDiffus}, addresses network traffic generation by transforming one-dimensional trace sequences into two-dimensional Gramian Angular Summation Field representations. These images capture features such as packet size and inter-arrival time, which are then used to train a diffusion model for synthesizing realistic traffic patterns.

Unlike their work, our model mainly focuses on generating user-level network traffic, as well as incorporating the trajectory data.

%% file: main/6.Conclusion.tex
\section{Conclusion}
In this work, we present a unified diffusion Transformer framework for joint simulation of mobile traffic and user trajectories. The model integrates wavelet-based multi-resolution decomposition, a multi-scale Transformer with cross-attention for capturing temporal dependencies, and a hybrid denoising mechanism that handles both continuous and discrete data. Semantic information is further incorporated via knowledge graph embeddings and a similarity-based transition mechanism for trajectory generation.

%% file: 0.main.bbl
\begin{thebibliography}{10}
\providecommand{\url}[1]{#1}
\csname url@samestyle\endcsname
\providecommand{\newblock}{\relax}
\providecommand{\bibinfo}[2]{#2}
\providecommand{\BIBentrySTDinterwordspacing}{\spaceskip=0pt\relax}
\providecommand{\BIBentryALTinterwordstretchfactor}{4}
\providecommand{\BIBentryALTinterwordspacing}{\spaceskip=\fontdimen2\font plus
\BIBentryALTinterwordstretchfactor\fontdimen3\font minus \fontdimen4\font\relax}
\providecommand{\BIBforeignlanguage}[2]{{%
\expandafter\ifx\csname l@#1\endcsname\relax
\typeout{** WARNING: IEEEtran.bst: No hyphenation pattern has been}%
\typeout{** loaded for the language `#1'. Using the pattern for}%
\typeout{** the default language instead.}%
\else
\language=\csname l@#1\endcsname
\fi
#2}}
\providecommand{\BIBdecl}{\relax}
\BIBdecl

\bibitem{ramamohanarao2017traffic}
K.~Ramamohanarao, J.~Qi, E.~Tanin, and S.~Motallebi, ``From how to where: Traffic optimization in the era of automated vehicles,'' in \emph{Proceedings of the 25th ACM SIGSPATIAL International Conference on Advances in Geographic Information Systems}, 2017, pp. 1--4.

\bibitem{24OpenData}
H.~Chai, T.~Jiang, and L.~Yu, ``Diffusion model-based mobile traffic generation with open data for network planning and optimization,'' in \emph{Proceedings of the 30th ACM SIGKDD Conference on Knowledge Discovery and Data Mining}, ser. KDD '24.\hskip 1em plus 0.5em minus 0.4em\relax New York, NY, USA: Association for Computing Machinery, 2024, p. 4828–4838.

\bibitem{22image1}
R.~Rombach, A.~Blattmann, D.~Lorenz, P.~Esser, and B.~Ommer, ``High-resolution image synthesis with latent diffusion models,'' in \emph{Proceedings of the IEEE/CVF Conference on Computer Vision and Pattern Recognition (CVPR)}, June 2022, pp. 10\,684--10\,695.

\bibitem{22image2}
C.~Saharia, W.~Chan, H.~Chang, C.~Lee, J.~Ho, T.~Salimans, D.~Fleet, and M.~Norouzi, ``Palette: Image-to-image diffusion models,'' in \emph{ACM SIGGRAPH 2022 Conference Proceedings}, 2022.

\bibitem{22VideoDiff}
\BIBentryALTinterwordspacing
J.~Ho, T.~Salimans, A.~Gritsenko, W.~Chan, M.~Norouzi, and D.~J. Fleet, ``Video diffusion models,'' in \emph{Advances in Neural Information Processing Systems}, S.~Koyejo, S.~Mohamed, A.~Agarwal, D.~Belgrave, K.~Cho, and A.~Oh, Eds., vol.~35.\hskip 1em plus 0.5em minus 0.4em\relax Curran Associates, Inc., 2022, pp. 8633--8646. [Online]. Available: \url{https://proceedings.neurips.cc/paper_files/paper/2022/file/39235c56aef13fb05a6adc95eb9d8d66-Paper-Conference.pdf}
\BIBentrySTDinterwordspacing

\bibitem{21CSDI}
Y.~Tashiro, J.~Song, Y.~Song, and S.~Ermon, ``Csdi: Conditional score-based diffusion models for probabilistic time series imputation,'' \emph{Advances in Neural Information Processing Systems}, vol.~34, pp. 24\,804--24\,816, 2021.

\bibitem{23DiT}
W.~Peebles and S.~Xie, ``Scalable diffusion models with transformers,'' in \emph{Proceedings of the IEEE/CVF International Conference on Computer Vision (ICCV)}, October 2023, pp. 4195--4205.

\bibitem{23DiffTraj}
Y.~Zhu, Y.~Ye, S.~Zhang, X.~Zhao, and J.~Yu, ``Difftraj: Generating gps trajectory with diffusion probabilistic model,'' in \emph{Advances in Neural Information Processing Systems}, A.~Oh, T.~Naumann, A.~Globerson, K.~Saenko, M.~Hardt, and S.~Levine, Eds., vol.~36.\hskip 1em plus 0.5em minus 0.4em\relax Curran Associates, Inc., 2023, pp. 65\,168--65\,188.

\bibitem{24ConTraj}
Y.~Zhu, J.~J. Yu, X.~Zhao, Q.~Liu, Y.~Ye, W.~Chen, Z.~Zhang, X.~Wei, and Y.~Liang, ``Controltraj: Controllable trajectory generation with topology-constrained diffusion model,'' in \emph{Proceedings of the 30th ACM SIGKDD Conference on Knowledge Discovery and Data Mining}, ser. KDD '24.\hskip 1em plus 0.5em minus 0.4em\relax New York, NY, USA: Association for Computing Machinery, 2024, p. 4676–4687.

\bibitem{liu2025traffic}
Z.~Liu, Q.~Long, H.~Wang, and Y.~Li, ``Multi-scale diffusion transformer for jointly simulating user mobility and mobile traffic pattern,'' in \emph{Proceedings of the 33rd ACM International Conference on Advances in Geographic Information Systems (SIGSPATIAL)}, 2025, pp. 1--4.

\bibitem{DDPM}
J.~Ho, A.~Jain, and P.~Abbeel, ``Denoising diffusion probabilistic models,'' \emph{CoRR}, vol. abs/2006.11239, 2020.

\bibitem{22Argmax}
E.~Hoogeboom, D.~Nielsen, P.~Jaini, P.~Forr\'{e}, and M.~Welling, ``Argmax flows and multinomial diffusion: Learning categorical distributions,'' in \emph{Advances in Neural Information Processing Systems}, M.~Ranzato, A.~Beygelzimer, Y.~Dauphin, P.~Liang, and J.~W. Vaughan, Eds., vol.~34.\hskip 1em plus 0.5em minus 0.4em\relax Curran Associates, Inc., 2021, pp. 12\,454--12\,465.

\bibitem{D3PM}
J.~Austin, D.~D. Johnson, J.~Ho, D.~Tarlow, and R.~van~den Berg, ``Structured denoising diffusion models in discrete state-spaces,'' in \emph{Advances in Neural Information Processing Systems}, M.~Ranzato, A.~Beygelzimer, Y.~Dauphin, P.~Liang, and J.~W. Vaughan, Eds., vol.~34.\hskip 1em plus 0.5em minus 0.4em\relax Curran Associates, Inc., 2021, pp. 17\,981--17\,993.

\bibitem{Wang15Traf_pattern}
H.~Wang, F.~Xu, Y.~Li, P.~Zhang, and D.~Jin, ``Understanding mobile traffic patterns of large scale cellular towers in urban environment,'' in \emph{Proceedings of the 2015 Internet Measurement Conference}, 2015, p. 225–238.

\bibitem{Smartphone}
T.~Li, T.~Xia, H.~Wang, Z.~Tu, S.~Tarkoma, Z.~Han, and P.~Hui, ``Smartphone app usage analysis: Datasets, methods, and applications,'' \emph{IEEE Communications Surveys \& Tutorials}, vol.~24, no.~2, pp. 937--966, 2022.

\bibitem{Decom}
B.~Cici, M.~Gjoka, A.~Markopoulou, and C.~T. Butts, ``On the decomposition of cell phone activity patterns and their connection with urban ecology,'' in \emph{Proceedings of the 16th ACM International Symposium on Mobile Ad Hoc Networking and Computing}, ser. MobiHoc '15, 2015, p. 317–326.

\bibitem{TuckER}
I.~Balazevic, C.~Allen, and T.~M. Hospedales, ``Tucker: Tensor factorization for knowledge graph completion,'' \emph{CoRR}, vol. abs/1901.09590, 2019.

\bibitem{ADAPTIVE}
S.~Zhang, T.~Li, S.~Hui, G.~Li, Y.~Liang, L.~Yu, D.~Jin, and Y.~Li, ``Deep transfer learning for city-scale cellular traffic generation through urban knowledge graph,'' 2023, p. 4842–4851.

\bibitem{24MSH-GAN}
T.~Li, S.~Hui, S.~Zhang, H.~Wang, Y.~Zhang, P.~Hui, D.~Jin, and Y.~Li, ``Mobile user traffic generation via multi-scale hierarchical gan,'' \emph{ACM Trans. Knowl. Discov. Data}, vol.~18, no.~8, Jul. 2024.

\bibitem{jiang2016timegeo}
S.~Jiang, Y.~Yang, S.~Gupta, D.~Veneziano, S.~Athavale, and M.~C. Gonz{\'a}lez, ``The timegeo modeling framework for urban mobility without travel surveys,'' \emph{Proceedings of the National Academy of Sciences}, vol. 113, no.~37, pp. E5370--E5378, 2016.

\bibitem{wang2023pategail}
H.~Wang, C.~Gao, Y.~Wu, D.~Jin, L.~Yao, and Y.~Li, ``Pategail: A privacy-preserving mobility trajectory generator with imitation learning,'' in \emph{Proceedings of the AAAI conference on artificial intelligence}, vol.~37, no.~12, 2023, pp. 14\,539--14\,547.

\bibitem{long2023practical}
Q.~Long, H.~Wang, T.~Li, L.~Huang, K.~Wang, Q.~Wu, G.~Li, Y.~Liang, L.~Yu, and Y.~Li, ``Practical synthetic human trajectories generation based on variational point processes,'' in \emph{Proceedings of the 29th ACM SIGKDD Conference on Knowledge Discovery and Data Mining}, 2023, pp. 4561--4571.

\bibitem{brockmann2006scaling}
D.~Brockmann, L.~Hufnagel, and T.~Geisel, ``The scaling laws of human travel,'' \emph{Nature}, vol. 439, no. 7075, pp. 462--465, 2006.

\bibitem{song2010modelling}
C.~Song, T.~Koren, P.~Wang, and A.-L. Barabási, ``Modelling the scaling properties of human mobility,'' \emph{Nature Physics}, vol.~6, no.~10, pp. 818--823, 2010.

\bibitem{jiawei2024large}
W.~JIAWEI, R.~Jiang, C.~Yang, Z.~Wu, R.~Shibasaki, N.~Koshizuka, C.~Xiao \emph{et~al.}, ``Large language models as urban residents: An llm agent framework for personal mobility generation,'' \emph{Advances in Neural Information Processing Systems}, vol.~37, pp. 124\,547--124\,574, 2024.

\bibitem{04Harpoon}
J.~Sommers and P.~Barford, ``Self-configuring network traffic generation,'' in \emph{Proceedings of the 4th ACM SIGCOMM Conference on Internet Measurement}, 2004, p. 68–81.

\bibitem{06Tmix}
M.~C. Weigle, P.~Adurthi, F.~Hern\'{a}ndez-Campos, K.~Jeffay, and F.~D. Smith, ``Tmix: a tool for generating realistic tcp application workloads in ns-2,'' vol.~36, no.~3, p. 65–76, jul 2006.

\bibitem{09Swing}
K.~V. Vishwanath and A.~Vahdat, ``Swing: Realistic and responsive network traffic generation,'' \emph{IEEE/ACM Transactions on Networking}, vol.~17, no.~3, pp. 712--725, 2009.

\bibitem{21Markov}
G.~Aceto, G.~Bovenzi, D.~Ciuonzo, A.~Montieri, V.~Persico, and A.~Pescapé, ``Characterization and prediction of mobile-app traffic using markov modeling,'' \emph{IEEE Transactions on Network and Service Management}, vol.~18, no.~1, pp. 907--925, 2021.

\bibitem{18DCNN}
C.~Zhang, H.~Zhang, D.~Yuan, and M.~Zhang, ``Citywide cellular traffic prediction based on densely connected convolutional neural networks,'' \emph{IEEE Communications Letters}, vol.~22, no.~8, pp. 1656--1659, 2018.

\bibitem{18LSTM}
A.~Dalgkitsis, M.~Louta, and G.~T. Karetsos, ``Traffic forecasting in cellular networks using the lstm rnn,'' in \emph{Proceedings of the 22nd Pan-Hellenic Conference on Informatics}, 2018, p. 28–33.

\bibitem{18LSTM2}
H.~D. Trinh, L.~Giupponi, and P.~Dini, ``Mobile traffic prediction from raw data using lstm networks,'' in \emph{2018 IEEE 29th Annual International Symposium on Personal, Indoor and Mobile Radio Communications (PIMRC)}, 2018, pp. 1827--1832.

\bibitem{24Atten}
J.~Su, H.~Cai, Z.~Sheng, A.~Liu, and A.~Baz, ``Traffic prediction for 5g: A deep learning approach based on lightweight hybrid attention networks,'' \emph{Digital Signal Processing}, vol. 146, p. 104359, 2024.

\bibitem{17FlowSyn}
S.~Iannucci, H.~A. Kholidy, A.~D. Ghimire, R.~Jia, S.~Abdelwahed, and I.~Banicescu, ``A comparison of graph-based synthetic data generators for benchmarking next-generation intrusion detection systems,'' in \emph{2017 IEEE International Conference on Cluster Computing (CLUSTER)}, 2017, pp. 278--289.

\bibitem{19Flow}
M.~Ring, D.~Schlör, D.~Landes, and A.~Hotho, ``Flow-based network traffic generation using generative adversarial networks,'' \emph{Computers \& Security}, vol.~82, pp. 156--172, 2019.

\bibitem{19PAC}
A.~Cheng, ``Pac-gan: Packet generation of network traffic using generative adversarial networks,'' in \emph{2019 IEEE 10th Annual Information Technology, Electronics and Mobile Communication Conference (IEMCON)}, 2019, pp. 0728--0734.

\bibitem{22Prac}
Y.~Yin, Z.~Lin, M.~Jin, G.~Fanti, and V.~Sekar, ``Practical gan-based synthetic ip header trace generation using netshare,'' in \emph{Proceedings of the ACM SIGCOMM 2022 Conference}.\hskip 1em plus 0.5em minus 0.4em\relax New York, NY, USA: Association for Computing Machinery, 2022, p. 458–472.

\bibitem{PcapGAN}
B.~Dowoo, Y.~Jung, and C.~Choi, ``Pcapgan: Packet capture file generator by style-based generative adversarial networks,'' in \emph{2019 18th IEEE International Conference On Machine Learning And Applications (ICMLA)}, 2019, pp. 1149--1154.

\bibitem{KE-GAN}
S.~Hui, H.~Wang, T.~Li, X.~Yang, X.~Wang, J.~Feng, L.~Zhu, C.~Deng, P.~Hui, D.~Jin, and Y.~Li, ``Large-scale urban cellular traffic generation via knowledge-enhanced gans with multi-periodic patterns,'' in \emph{Proceedings of the 29th ACM SIGKDD Conference on Knowledge Discovery and Data Mining}, 2023, p. 4195–4206.

\bibitem{20traffic}
S.-S. Kim, M.~Chung, and Y.-K. Kim, ``Urban traffic prediction using congestion diffusion model,'' in \emph{2020 IEEE International Conference on Consumer Electronics - Asia (ICCE-Asia)}, 2020, pp. 1--4.

\bibitem{Qi2024CANDLE}
X.~Qi, H.~Chai, L.~Yu, Y.~Li, and Z.~Wang, ``Regional features conditioned diffusion models for 5g network traffic generation,'' in \emph{Proceedings of the 32nd ACM International Conference on Advances in Geographic Information Systems}, ser. SIGSPATIAL '24.\hskip 1em plus 0.5em minus 0.4em\relax New York, NY, USA: Association for Computing Machinery, 2024, p. 396–409.

\bibitem{Chai25STK}
H.~Chai, X.~Qi, and Y.~Li, ``Spatio-temporal knowledge driven diffusion model for mobile traffic generation,'' \emph{IEEE Transactions on Mobile Computing}, pp. 1--18, 2025.

\bibitem{24NetDiffus}
N.~Sivaroopan, D.~Bandara, C.~Madarasingha, G.~Jourjon, A.~P. Jayasumana, and K.~Thilakarathna, ``Netdiffus: Network traffic generation by diffusion models through time-series imaging,'' \emph{Computer Networks}, vol. 251, p. 110616, 2024.

\end{thebibliography}
